\documentstyle[12pt]{article}
\def\sign{{\rm sign}\;}
\def\const{{\rm const}}

\def\e{{\rm e}}
\def\sumii{\sum_{i=2}^{n}}
\def\sumi{\sum_{i=1}^{n}}
\def\RR{\overline{R}}
\def\ds{d\overline{s}^2}
\def\half{\frac{1}{2}}
\def\to{\rightarrow}
\def\then{\Rightarrow}
\def\gg{\overline{g}}
\def\nn{\nonumber\\}
\newcommand{\vars}[1]{\left\{\begin{array}{ll}#1\end{array}\right.}
\newcommand{\wt}[1]{\widetilde{#1}}

\begin{document}
\thispagestyle{empty}
\setcounter{page}{0}

\begin{center}
               RUSSIAN GRAVITATIONAL SOCIETY\\
               CENTER FOR SURFACE AND VACUUM RESEARCH\\
               DEPARTMENT OF FUNDAMENTAL INTERACTIONS AND METROLOGY\\
\end{center}
\vskip 4ex
\begin{flushright}                 RGS-CSVR-013/94\\
                                   gr-qc/9410038
\end{flushright}
\vskip 15mm

\begin{center}
{\large\bf VACUUM WEYL COSMOLOGIES IN D DIMENSIONS}\\
\vskip2.5ex
     {\bf K.A.Bronnikov and V.N.Melnikov}\\
\vskip 5mm
     {\em Center for Surface and Vacuum Research,\\
     8 Kravchenko str., Moscow 117331, Russia}\\
     e-mail: bron@cvsi.uucp.free.net\\

\vskip 10mm
\end{center}

{\bf ABSTRACT}

\bigskip
\noindent
        Vacuum cosmological models are considered in the context of a
        multidimensional theory of gravity with integrable Weyl geometry.
        A family of exact solutions with a chain of internal spaces
        is obtained. Models with one internal space are considered in
        more detail; nonsingular models are selected.

\vskip 65mm

\centerline{Moscow 1994}
\pagebreak

\setcounter{page}{1}

\bigskip
\noindent {\bf 1.}
     Scalar fields play a very significant role in modern cosmology
     \cite{hydro}, in particular, in various inflationary models. However,
     there is always a problem of the origin of this field. It can be
     naturally solved in multidimensional models (scalar fields are
     represented by extra-dimension scale factors \cite{bm-itogi}) and
     other models with generalized geometries such as the Weyl geometry
     \cite{nov}. In both case nonsingular cosmological models have been
     obtained.

     Here we would like to consider a scheme uniting the two approaches,
     i.e., multidimensional cosmology with an integrable Weyl geometry.
     Since the most reliable results are obtained on the basis of exact
     solutions, we try to find them in the simplest cases. Recently some
     results in the same scheme have been obtained using numerical methods
     \cite{konstmel}.

\bigskip
\noindent {\bf 2.} Consider a $D$-dimensional manifold $W_D$ with
     an integrable Weyl geometry defined by the metric $g_{AB}$ and the
     connection
\begin{equation}
     \Gamma^A_{BC}= \wt{\Gamma}^A_{BC}
     -\half(\omega_B\delta^A_C + \omega_C\delta^A_B-g_{BC}\omega^A)

           \label{Conn}
\end{equation}
        where $\wt{\Gamma}^A_{BC}$ are the Christoffel symbols for the
        metric $g_{AB}$, $\omega$ is a scalar field and
        $\omega_A = \partial_A\omega$.

        Thus the gravitational field is determined by the tensor $g_{AB}$ and
        the scalar $\omega$, just as in the scalar-tensor theories (STT) of
        gravity. As is the case with STT, in general the gravitational
        Lagrangian may contain various invariant combinations of $g_{AB}$ and
        $\omega$. Let us restrict ourselves to Lagrangians (a) linear in the
        scalar curvature and (b) quadratic in $\omega_A$. Then the general form
        of the Lagrangian is
\begin{equation}
        L = A(\omega)R + B(\omega)\omega^A\omega_A - 2\Lambda(\omega)+L_m

          \label{Lagr}
\end{equation}
        where $R$ is the Weyl scalar curvature corresponding to the connection
        (\ref{Conn}), $A,\ B$ and $\Lambda$ are arbitrary functions and $L_m$
is
        the nongravitational matter Lagrangian.

        To simplfy the field equations let us make use of the expression of $R$
        in terms of the Riemannian curvature  $\wt{R}$ corresponding to the
        metric $g_{AB}$
\begin{equation}
        R = \wt{R}+ (D-1)\wt{\Box}\omega
                -\frac{1}{4}(D-1)(D-2)\omega^A\omega_A        \label{Rtransit}
\end{equation}
        (in $\wt{R}$ and $\wt{\Box}$ the Riemannian connection
        $\wt{\Gamma}^A_{BC}$ is used) and the conformal mapping well-known in
     STT \cite{wag}, modified for $D$ dimensions \cite{birk}:
\begin{equation}
        g_{MN}= A^{-2/(D{-}2)}\gg_{MN}.
\label{Conf}
\end{equation}
        Consequently, omitting a total divergence, we obtain the following form
        of the Lagrangian:
\begin{equation}
        \overline{L} = A(\omega)\overline{R} +
        F(\omega)\gg_{AB}\omega_A\omega_B
        + A^{-D/(D{-}2)}[- 2\Lambda(\omega)+L_m]

          \label{Lagr1}
\end{equation}
        where $\overline{R}$ is the Riemannian scalar curvature corresponding
to
        the metric $\gg_{AB}$ and ($A_\omega\equiv dA/d\omega$)
\begin{equation}
        F(\omega) = \frac{1}{A^2}\biggl[AB-
                (D{-}1)A\biggl(A_\omega+\frac{D{-}2}{4}\biggr)+
                \frac{D{-}1}{D{-}2}A_\omega^2 \biggr].
\label{DefF}
\end{equation}

\bigskip
\noindent{\bf 3.} Let us consider vacuum cosmological models in the
     theory described: put $L_m=0$ and postulate the following structure of
        the space-time $W_D$:
\begin{equation}
        W_D = {\rm R}\times M_1 \times \ldots \times M_n;\qquad
                \dim M_i = N_i;
\label{Stru}
\end{equation}
        the subspaces $M_i$ are assumed to be maximally symmetric. The
component
        \ R\ corresponds to the time $\tau$; besides, we assume
        $\omega=\omega(\tau)$. Thus the effective Riemannian metric is written
        in the form
\begin{equation}
     \ds = \gg_{AB}dx^A dx^B = \e^{2\gamma(\tau)}d\tau^2-
                 \sumi \e^{2\beta_i(\tau)} ds_i^2
\label{Ds}
\end{equation}
        where $ds_i^2$ are $\tau$-independent metrics of the $N_i$-dimensional
        spaces of constant curvatures $K_i$; with no loss of generality one can
        put $K_i = 0,\ \pm 1$.

     Making use of the freedom to choose the time coordinate $\tau$, let us
     introduce the harmonic time by putting
\begin{equation}
        \gamma= \sumi N_i \beta_i.
\label{Harm}
\end{equation}
        Then the Ricci tensor for $\gg_{AB}$ has the following nonzero
        components:
\begin{eqnarray}
   \RR^\tau_\tau &=& \e^{-2\gamma}
     \Bigl(\ddot{\gamma}-\dot{\gamma}^2+ \sumi N_i\dot{\beta}_i^2\Bigr), \nn
   \RR^{m_i}_{n_i} &=& \delta^{m_i}_{n_i}
     \Bigl[\e^{-2\gamma}\ddot{\beta}_i+ (N_i{-}1)K_i\e^{-2\beta_i}\Bigr]

           \label{Ricci}
\end{eqnarray}
        where the indices $m_i,\ n_i$ belong to the subspace $M_i$.

\bigskip
\noindent{\bf 4.}
        The field equations take an especially simple form under the additional
        condition $\Lambda\equiv 0$:
\begin{eqnarray}
        \RR_{MN} + F(\omega)\omega_M \omega_N =0,
\label{Einstein}\\
        2\overline{\nabla}_M[F(\omega)\omega^M] -
                F_{\omega}\omega^M\omega_M =0.                  \label{Eomega}
\end{eqnarray}
     They can be integrated completely under the above assumptions if
     (i) all the subspaces $M_i$ are Ricci-flat and (ii) if
     one of $M_i$ (for instance, $M_1$) is a space of nonzero constant
        curvature ($K_1$). Indeed, putting $K_i=0\ (i>1)$, we obtain:
\begin{eqnarray}
        (F\dot{\omega}^2)^. =0 &\then& F\dot{\omega}^2 = \const;

        \label{Omdot}\\
        \ddot{\beta}_i =0      &\then& \beta_i=\beta_{i0}+h_i\tau,
                                \qquad i>1;
\label{Betai}\\
        \ddot{\gamma}-\ddot{\beta}_1 &=&-K_1 d^2 \e^{2\gamma-2\beta_1}

     \label{Ebeta1}
\end{eqnarray}
        where $d{+}1 = N_1 = \dim M_1$. The equation (\ref{Ebeta1}) leads to
     different results for different $K_1$: for $K_1=0$
     (case (i)) Eq.(\ref{Betai}) may be regarded to include $i=1$; for
     $K_1\ne 0$ (case (ii)) we get:
\begin{eqnarray}
        \e^{\beta_1-\gamma} &=& \frac{d}{k} \cosh k\tau, \qquad k>0
                        \qquad (K_1=+1),
\label{Beta1+}\\
        \e^{\beta_1-\gamma} &=& d\cdot s(k,\tau)\equiv
                \vars{ (d/k) \sinh k\tau, \qquad & k>0,\\
                                   d\cdot\tau, & k=0,\\
        (d/k) \sin k\tau, & k<0, } \qquad( K_1=-1)      \label{Beta1-}
\end{eqnarray}
        where $k=\const$ and another integration constant is elimintaed by a
        particular choice of the origin of $\tau$. Lastly, a combination of
        components of (\ref{Einstein}) representing the temporal component
     of the Einstein equations (the initial data equation) leads to the
        following relation among the integration constants:
\begin{eqnarray}
        \biggl(\sumi N_i h_i\biggr)^2-\sumi N_i h_i^2=S,   &&\ K_1=0;

        \label{Int0}\\
        \frac{d{+}1}{d}k^2\sign k=
        \frac{1}{d}\biggl( \sumii N_i h_i\biggr)^2
                + \sumii N_i h_i^2 + S,      && K_1\ne 0.    \label{Int1}
\end{eqnarray}

        Thus the set of equations (\ref{Einstein}),(\ref{Eomega}) has been
        integrated in quadratures.

     As the original functions $A(\omega)$ and $B(\omega)$ and hence
     $F(\omega)$ are arbitrary, it is difficult to describe the physical
     properties of the models in a general form. Therefore here we would like
     to restrict ourselves to some simple special cases.

     Thus, we will assume $A\equiv 1$ while $B(\omega)$ remains arbitrary, so
     that the metrics $\gg_{AB}$ and $g_{AB}$ coincide.

\bigskip
\noindent
     {\bf 5.} As the first step consider 4-dimensional cosmologies: put
     $n=1,\ d=2,\ \beta_1\equiv \beta(\tau)$. The condition that $\tau$ is a
     harmonic coordinate takes the form $\gamma=3\beta$ and for the scale
     factor we get:
\begin{equation}
     \e^{2\beta}= a^2(\tau) =\vars{
                         1/2s(k,\tau),\ \ & K_1=1,\\
                         \e^{k\tau},      & K_1=0,\\
                         1/2\cosh k\tau,\ \ & K_1=-1,     } \label{a1}
\end{equation}
     while the physical time is determined by the integral
     $t= \pm\int \e^{\gamma(\tau)}d\tau$. The constant $k$ is connected with
     the ``scalar charge'' $S$ according to (\ref{Int0}), (\ref{Int1}) where
     one should substitute $h_i=0\ (i>1)$ and $h_1=k/2$:
\begin{equation}
     2S=\vars{    3k^2\sign k,\ \ & K_1 = \pm 1,\\
                  3k^2,           & K_1 =0.}                \label{Int2}
\end{equation}

     It is easy to obtain that in the case of a spherical world ($K_1=1$) the
     values $\tau=\pm\infty$ correspond to finite times $t_1$ and $t_2$ at
     which $a=0$ (the initial and final singularities). For a flat world
     ($K_1=0$) at $k\ne 0$ and a hyperbolic one ($K_1=-1$) at $k>0$ an
        initial or final singularity is observed at infinite $\tau$. In the
        special case $K_1=-1,\ k=0$ we obtain the Milne vacuum model which is
        known to describe a domain in flat space-time (in this case $S=0$, so
        that the scalar field is trivial).

     Lastly, in the case $K_1=-1,\ k<0$ we see that the limits
     $\tau\to 0,\ \pi/|k|$ correspond to $t\to\pm\infty$; the scale factor
     $a(t)$ decreases in an asymptotically linear manner in the remote past
     ($t\to -\infty$), reaches a minimum at $\tau=\pi/2|k|$ and grows in an
     asymptotically linear manner at $t\to\infty$. The model is
     time-symmetric with respect to the maximum contraction instant.

     By (\ref{Int2}) a necessary condition for the existence of nonsingular
     solutions is the restriction $F<0$ on the function (\ref{DefF}), or,
     in terms of the initial function $B(\omega):\ B<3/2$.

     These results confirm those of Ref.  \cite{nov}.

\bigskip
\noindent
     {\bf 6.} Consider now the metric $\gg_{AB}$ for $n=2$: let
     $a(t)\equiv\e^{\beta_1(\tau)}$ be the scale factor of the ordinary
     physical space ($N_1=3$), while $b(t)\equiv \e^{\beta_2(\tau)}$ that of
     the internal space ($N_2 = N$).

\medskip
\noindent
     {\bf 6.1.} In the case $K_1=0$ (spatially flat models) we obtain:
\begin{equation}
     \ds = \e^{2(3h_1+Nh_2)\tau}d\tau^2 - \e^{2h_1\tau}ds_1^2
               - \e^{2h_2\tau}ds_2^2                             \label{Ds2}
\end{equation}
     where with no loss of generality the scales in $M_1$ and  $M_2$ are
        chosen so that $\beta_{10}=\beta_{20}=0$. Herewith
\begin{equation}
     6(h_1 + Nh_2/2)^2 = N(N+1/2) +S                             \label{Int3}
\end{equation}

     In the special case $3h_1+Nh_2 =0$ the time coordinate $\tau$ is
     synchronous, in other words, physical. The metric (\ref{Ds2}) is
     nonsingular at finite $\tau$ and describes an exponential expansion
     (inflation) of one of the spaces (e.g., the physical one, $M_1$) and a
     simultaneous exponential contraction of the other, $M_2$, since $h_1$
     and $h_2$ have different signs. However, by (\ref{Int3}) and
     (\ref{Omdot})
\begin{equation}
     S=F\dot{\omega}^2 = -h_1^2(2N+1)/N <0.                     \label{SF}
\end{equation}
     So a necessary condition for the existence of the special solution
     (\ref{Ds2}) is the restriction
\begin{equation}
     B(\omega) < (D-1)(D-2)/4,                                 \label{B-less}
\end{equation}
     more general than $B<3/2$ from Sect. 4.

     In the more general case $3h_1 + Nh_2 =H\ne 0$ a transition to the
     physical time $dt = \e^{H\tau}d\tau$ leads to the metric
\begin{equation}
     \ds = dt^2 - t^{2h_1/H}ds_1^2 - t^{2h_2/H}ds_2^2           \label{Ds3}
\end{equation}
     which is singular at $t=0$ if at least one of the constants $h_1$ or
     $h_2$ is nonzero. At $h_1=h_2=0$ the metric is static and (\ref{SF})
     implies that either $\dot{\omega}=0$ (the solution is trivial), or
     $F\equiv 0$, a special choice of $B$ such that $\omega(\tau_)$ has no
     dynamics.

\medskip
\noindent
     {\bf 6.2.} For a spherical world ($K_1=1$) the metric is
\begin{equation}
     \ds= \frac{\e^{-Nh\tau}}{2\cosh k\tau}\biggl[
          \frac{d\tau^2}{4\cosh^2 k\tau} - ds_1^2\biggr] -\e^{2h\tau}ds_2^2
                                                               \label{Ds4}
\end{equation}
     where $ds_1^2$ is the line element on a unit sphere. A consideration
     like that in Sect. 5.1 leads to the following conclusions:
\begin{description}
\item[(a)] The model behavior is classified by the values of the constant
     $h=h_2$ as compared with $k>0$. The physical time
     $t= \pm\int\e^{\gamma(\tau)}d\tau$ varies either within a finite segment
     $[t_1,\ t_2]$ (if $|Nh|< 3k$), or within a semi-infinite range (if
     $|Nh|\geq 3k$).
\item[(b)] At any finite boundary of the range of $t$ at least one of the
     scale factors $a(t)$ or $b(t)$ vanishes, i.e., a singularity takes place.
\item[(c)] At $t\to \pm\infty$ either $a\to 0,\ b\to\infty$, or conversely,
     $a\to\infty,\ b\to 0$.
\end{description}
     The value $S=F\dot{\omega}^2$ is determined at $K_1= \pm 1$ from
\begin{equation}
     3k^2\sign k =N(N+2)h^2 +2S.                             \label{Int4}
\end{equation}

\medskip
\noindent
     {\bf 6.3.} For hyperbolic models ($K_1=-1$) the metric has the form
\begin{equation}
     \ds= \frac{\e^{-Nh\tau}}{2s(k,\tau)}\biggl[
          \frac{d\tau^2}{4s^2(k,\tau)} - ds_1^2\biggr] -\e^{2h\tau}ds_2^2
                                                               \label{Ds5}
\end{equation}
     (the same as (\ref{Ds4}) but the function $\cosh k\tau$ is replaced by
     $s(k,\tau)$ defined in (\ref{Beta1-})). Preserving generality, let us
     assume $\tau >0$.

     The model behavior may be briefly described as follows:
\begin{description}
\item[(a)] At $k>0,\ Nh\leq -3k$ or $k=0,\ h<0$ the physical time
     $t= \pm\int\e^{\gamma(\tau)}d\tau$ ranges from $-\infty$ to $+\infty$.
     The factor $b(t)=\e^{h(\tau)}$ varies from a finite value at $\tau= 0$
     ($t=-\infty$) to zero at $\tau\to\infty\ (t\to\infty)$. The factor
     $a(t)$ describes a power-law contraction from infinity (at $t\to
     -\infty$) to a regular minimum and an infinite (in general, power-law)
     expansion at $t\to\infty$. There is no singularity at finite $t$.
\item[(b)] At $k\geq 0,\ Nh>3k$ the model is singular at finite $t$
     corresponding to $\tau\to\infty$. In the special case $h=k=0$ we come
     again to the Milne model (see Sect.4) supplemented with the space $M_2$
     with a constant scale factor.
\item[(c)] At $k<0$ the time $t$ ranges again from $-\infty$ to $+\infty$.
     The factor $a(t)$ behaves as it did in item (a), however, its variation
     at $t\to\pm\infty$ is linear (but in general with unequal
     slopes at the two asymptotics). The factor $b(t)$ changes monotonically
     between two finite boundary values.
\end{description}

     Unlike the 4-dimensional models (Sect. 4), the nonsingular
     multidimensional ones with $h\ne 0$ exhibit a time-asymmetric behavior
     of $a(t)$.

     It is seen in a straightforward way that in all the nonsingular models
     the requirement (\ref{B-less}) is imposed on $B(\omega)$, which, as it
     could be formulated in general relativity, means the negative scalar
     field energy density.

     Some properties of the above models have been discovered in numerical
     calculations for a number of special cases with $D=5$ and $D=6$
     \cite{konstmel}.

     In conclusion, we have seen that some of the multidimensional Weyl
     cosmologies are nonsingular: there are special flat-space models with
     eternally increasing or decreasing scale factors (such models are
     absent in the 4-dimensional approach) and there are more general
     hyperbolic models with a cosmological bounce (generalizing the
     4-dimensional ones \cite{nov}).  However, it should be taken into
     account that we have considered only one conformal gauge (although, in
     a certain sense, the most natural one), while in others the picture of
     singularities may change. The choice of a conformal gauge, connected
     with the choice of a system of measurements, is a separate problem
     \cite{hydro}, especially when a generalized geometry is used; its
     solution depends on the specific form of interaction between matter
     and geometry, out of the scope of vacuum cosmologies.

\end{document}